# Quantitative Benchmarking of Spectroscopic Homogeneous and Inhomogeneous Linewidth Separation

**Adam Alfrey,[1,2] Cole Tait,[1] Joshua R. Hendrickson,[2] and Steven T. Cundiff[1,3*]**

[1]*Department of Physics, University of Michigan, 450 Church St, Ann Arbor, Michigan 48109, USA*
[2]*Air Force Research Laboratory, 2241 Avionics Circle, Wright-Patterson Air Force Base, Ohio 45433, USA*
[3]*Quantum Research Institute, University of Michigan, Ann Arbor, Michigan 48109, USA*
*[cundiff@umich.edu](mailto:cundiff@umich.edu)

**Separating homogeneous and inhomogeneous contributions to a spectral linewidth is essential for connecting optical spectra to microscopic dephasing and disorder. We quantitatively compare linewidth separation by one-dimensional Voigt analysis and two-dimensional coherent spectroscopy (2DCS). Using photoluminescence and 2DCS measurements of the same hBN-encapsulated monolayer $MoSe_2$ sample, we construct uncertainty-weighted $\chi^2$ landscapes for the Gaussian inhomogeneous width and Lorentzian homogeneous half-width. Although a Voigt profile accurately reproduces the photoluminescence spectrum, the two widths are strongly anticorrelated and therefore poorly determined independently. By contrast, joint fitting of diagonal and cross-diagonal 2DCS slices produces more compact confidence regions with reduced parameter correlation and elongation. These results quantitatively demonstrate the improved linewidth separation provided by 2DCS and illustrate the limitations of assigning homogeneous and inhomogeneous contributions from one-dimensional Voigt fits alone.**

The width and shape of a spectral line reflect microscopic dissipation and disorder. A distribution of resonance frequencies produces inhomogeneous broadening, typically modeled by a Gaussian width $\sigma$, whereas homogeneous Markovian dephasing produces a Lorentzian half-width $\gamma$; their convolution is a Voigt profile [1]. Voigt fits are widely applied to one-dimensional spectra, with $\sigma$ and $\gamma$ interpreted as inhomogeneous and homogeneous contributions, respectively. However, agreement with a Voigt profile does not ensure that these parameters are independently determined: a one-dimensional spectrum may tightly constrain the overall linewidth while leaving $\sigma$ and $\gamma$ strongly anticorrelated.

Two-dimensional coherent spectroscopy (2DCS) provides complementary constraints because inhomogeneous broadening elongates the diagonal response, whereas homogeneous broadening primarily determines the cross-diagonal width [2–10]. The corresponding analytic lineshapes and joint-fitting strategy were established by Siemens *et al.* [11]. Here, we quantitatively compare homogeneous–inhomogeneous linewidth separation by one-dimensional Voigt analysis and joint fitting of diagonal and cross-diagonal 2DCS slices. We evaluate the two approaches using uncertainty-weighted $\chi^2$ landscapes, parameter correlations, and confidence-region elongations.

We demonstrate this comparison using the exciton resonance of hBN-encapsulated monolayer $MoSe_2$ at 8 K. Excitonic linewidths in transition-metal dichalcogenide monolayers encode coherence dynamics and disorder [12–21], while hBN encapsulation narrows these resonances toward the homogeneous limit [22,23]. The resulting regime, in which homogeneous and inhomogeneous contributions are comparable, provides a stringent test of their quantitative separation. We measure the same sample using photoluminescence (PL) and 2DCS and construct the full uncertainty-weighted $\chi^2(\sigma,\gamma)$ landscape for each method [24].

PL spectra of the hBN-encapsulated $MoSe_2$ flake are measured using a grating spectrometer and CCD under continuous-wave HeNe excitation (632.8 nm) focused to a ~ 1 $\mu$m spot. The excitonic PL spectrum is modeled as a Voigt profile

$$y(\omega) = AV(\omega - \omega_X; \sigma, \gamma) + B, \tag{1}$$

where $A$ is the peak amplitude, $\omega_X$ is the exciton center frequency, $B$ is a constant offset, and the Voigt profile, $V(\Delta;\sigma,\gamma)$, is the convolution

$$V(\Delta;\sigma,\gamma) = \int_{-\infty}^{\infty} G(\Delta - \omega';\sigma)\, L(\omega';\gamma)\, d\omega', \tag{2}$$

with $\Delta \equiv \omega - \omega_X$ and the normalized Gaussian and Lorentzian components are

$$G(\Delta;\sigma) = \frac{1}{\sqrt{2\pi}\sigma} \exp\left(-\frac{\Delta^2}{2\sigma^2}\right) \tag{3}$$

and

$$L(\Delta;\gamma) = \frac{1}{\pi}\frac{\gamma}{\Delta^2 + \gamma^2}. \tag{4}$$

The Gaussian width $\sigma$ parameterizes inhomogeneous broadening and the Lorentzian half-width $\gamma$ parameterizes homogeneous dephasing. Fig. 1 compares Gaussian, Lorentzian, and Voigt fits to the measured PL spectrum.

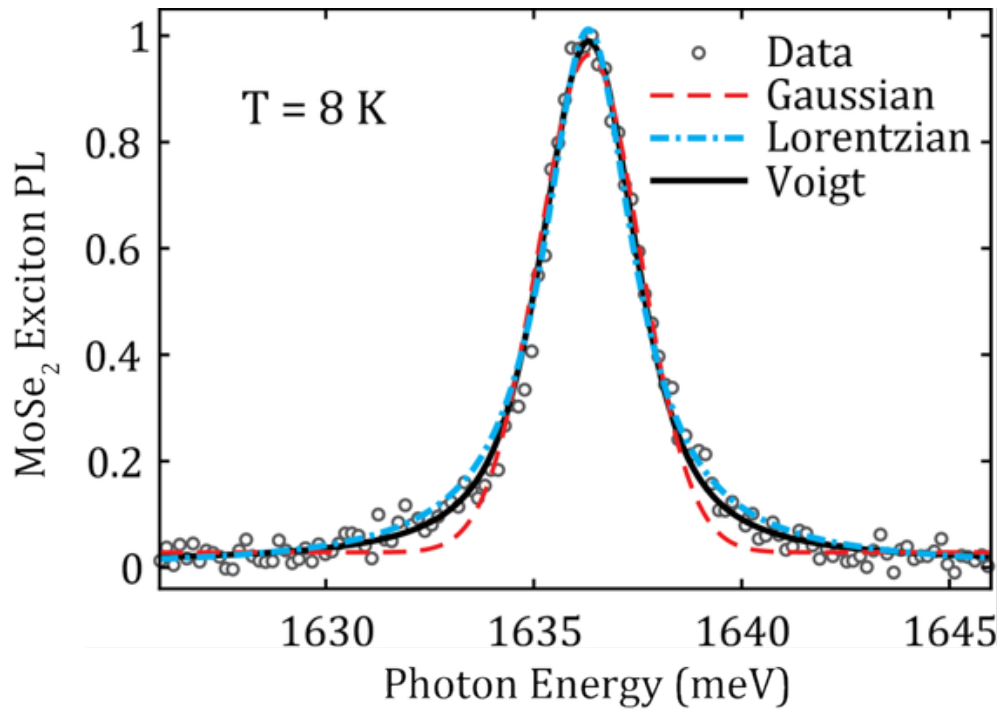

Fig. 1. **MoSe₂ exciton photoluminescence is accurately reproduced by a Voigt profile**. Normalized PL spectrum at 8 K (open circles), excited by a 632.8 nm HeNe laser at 20 μW average power, with best-fit Gaussian (red dashed), Lorentzian (blue dashed), and Voigt (black solid) profiles. The Voigt profile reproduces both the peaked core and extended wings more accurately than either single-component model.

To determine the count-dependent experimental uncertainty, we acquire $M = 5$ repeated calibration spectra at one sample location. For spectral point $i$, the mean count and sample variance are,

$$\bar{y}_i^{cal} = \frac{1}{M}\sum_{j=1}^{M} y_{i,j}^{cal}, \quad s_i^2 = \frac{1}{M-1}\sum_{j=1}^{M}\left(y_{i,j}^{\mathrm{cal}} - \bar{y}_i^{cal}\right)^2. \quad (5)$$

We fit the resulting variance-versus-mean-count relation with the linear model $s^2(\bar{y}) = \alpha\bar{y} + \beta$. For each spectrum analyzed below, the uncertainty at spectral point $i$ is then assigned from its measured count as $\epsilon_i = \sqrt{\alpha y_i + \beta}$ . This calibration is assumed to apply across the measured sample region. Each individual spectrum is fit with the Voigt model in Eq. (1) by minimizing the weighted sum of squared residuals

$$\chi^2_{\mathrm{PL}}(A, \omega_X, \sigma, \gamma, B) = \sum_{i=1}^{N}\left(\frac{y_i - [AV(\Delta_i; \sigma, \gamma) + B]}{\epsilon_i}\right)^2 \quad (6)$$

where $y_i$ is the single spectrum being fit. To quantify how well the one-dimensional PL spectrum independently constrains $\sigma$ and $\gamma$, we profile the $\chi^2$ surface by constructing a $(\sigma, \gamma)$ grid and redo the fit with respect to the parameters $(A, \omega_X, B)$ at each grid point,

$$\chi^2_{\mathrm{PL}}(\sigma, \gamma) = \min_{A,\omega_X,B} \chi^2_{\mathrm{PL}}(A, \omega_X, \sigma, \gamma, B). \quad (7)$$

Fig. 2 shows joint confidence regions from $\Delta\chi^2_{\mathrm{PL}}(\sigma, \gamma)$. Their elongation demonstrates that an increase in $\sigma$ can be compensated by a decrease in $\gamma$ with little change in $\chi^2$, and vice versa. Thus, although the Voigt profile accurately reproduces the measured spectrum, the individual contributions are not independently determined by the PL data alone.

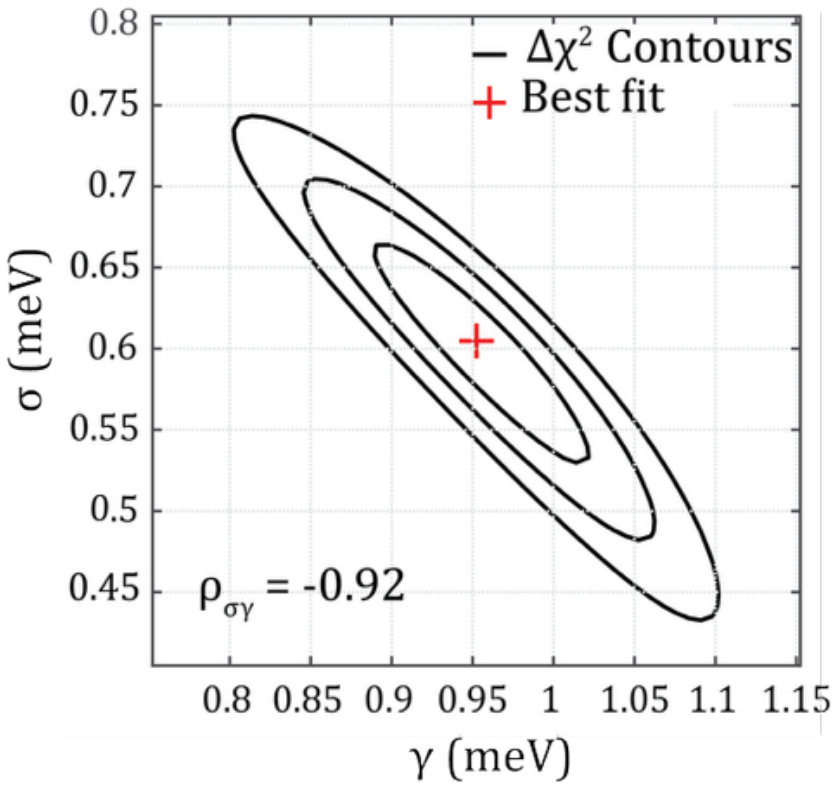

Fig. 2. **One-dimensional Voigt fitting leaves the linewidth decomposition poorly constrained**. Profiled $\Delta\chi^2_{\mathrm{PL}}(\sigma, \gamma)$ landscape for the spectrum in Fig. 1, obtained by scanning $(\sigma, \gamma)$ while optimizing $(A, \omega_X, B)$. Contours show the 68.27%, 95.45%, and 99.73% joint confidence regions for two parameters; the red + marks the best fit. Their elongation shows that compensating changes in $\sigma$ and $\gamma$ reproduce the one-dimensional spectrum nearly equally well.

We next analyze two orthogonal projections from a measured two-dimensional coherent spectrum of the same MoSe₂ exciton. 2DCS resolves the nonlinear optical signal as a function of two frequency axes; often designated as the excitation and detection frequencies [6,8,11,17]. Inhomogeneous broadening elongates the response along the diagonal (where excitation and detection energies are correlated), while homogeneous dephasing sets the cross-diagonal width, providing the orthogonal sensitivity to $\sigma$ and $\gamma$ that one-dimensional spectra lack [2,11,25,26].

Measurements are performed with a MONSTR Sense BigFoot nonlinear spectrometer. Ultrafast pulses (~100 fs) from an 80 MHz Ti:sapphire mode-locked oscillator are split, frequency-tagged by acousto-optic modulators, and routed through nested Michelson interferometers that provide phase-stable inter-pulse delays for $t$, $\tau$, and $T$ [6]. The nonlinear signal is isolated by lock-in detection at the appropriate modulation frequency and recorded as a function of $t$ and $\tau$; 2D Fourier transformation then yields the 2D spectrum shown in Fig. 3(a). All spectra were acquired at an average power of 1 $\mu$W per beam at the sample; power-dependent measurements confirmed operation in the $\chi^{(3)}$ regime. The spectra are not corrected to account for the finite laser bandwidth as the laser spectrum, shown at the top of Fig. 3(a), is much broader than the resonance and thus will cause minimal distortion. Both slices in Figures 3(b) and 3(c) are expressed as detunings from the peak center frequency $\omega_X$. We model the diagonal and cross-diagonal lineshapes using the analytic expressions derived by Siemens et al. [11]. The cross-diagonal slice, which is predominantly sensitive to homogeneous dephasing, is

$$S_{\mathrm{cross}}(\omega_{t'}; \sigma, \gamma) = \left| \frac{\exp\left(\frac{(\gamma - i\omega_{t'})^2}{2\sigma^2}\right) \mathrm{erfc}\left(\frac{\gamma - i\omega_{t'}}{\sqrt{2}\sigma}\right)}{\sigma(\gamma - i\omega_{t'})} \right|, \quad (8)$$

while the diagonal slice, which is sensitive to the energetic distribution of optically addressed transitions within the laser spot is

$$S_{\text{diag}}(\omega_{\tau'};\sigma,\gamma) = \frac{2\pi}{\gamma}\text{Voigt}(\omega_{\tau'};\sigma,\gamma), \tag{9}$$

where erfc is the complementary error function and Voigt is defined in Eq. (2). Crucially, these two projections share the same physical parameters $(\sigma,\gamma)$ but depend on them through different functional forms: the cross-diagonal width is governed primarily by $\gamma$, whereas the diagonal width reflects the full Voigt convolution of $\sigma$ and $\gamma$. This complementary sensitivity enables improved separability when the two slices are fit jointly. We construct a joint objective function that shares $(\sigma,\gamma)$ across both slices while allowing each an independent amplitude and offset,

$$\chi^2_{\text{2DCS}}(\sigma,\gamma) = \min_{p}\sum_{i=1}^{N}\left[\left(\frac{y_i^C - (A_C S_{\text{cross}}(\omega_i;\sigma,\gamma) + B_C)}{\epsilon_i^C}\right)^2 + \left(\frac{y_i^D - (A_D S_{\text{diag}}(\omega_i;\sigma,\gamma) + B_D)}{\epsilon_i^D}\right)^2\right], \tag{10}$$

where $y_i^C$ and $y_i^D$ are the cross-diagonal and diagonal slices, respectively; $p = (A_C, B_C, A_D, B_D)$ collects their independent amplitudes and offsets; and $\epsilon_i^C$ and $\epsilon_i^D$ are the corresponding per-pixel uncertainties. The per-pixel uncertainties are assigned using the count-dependent calibration described above.

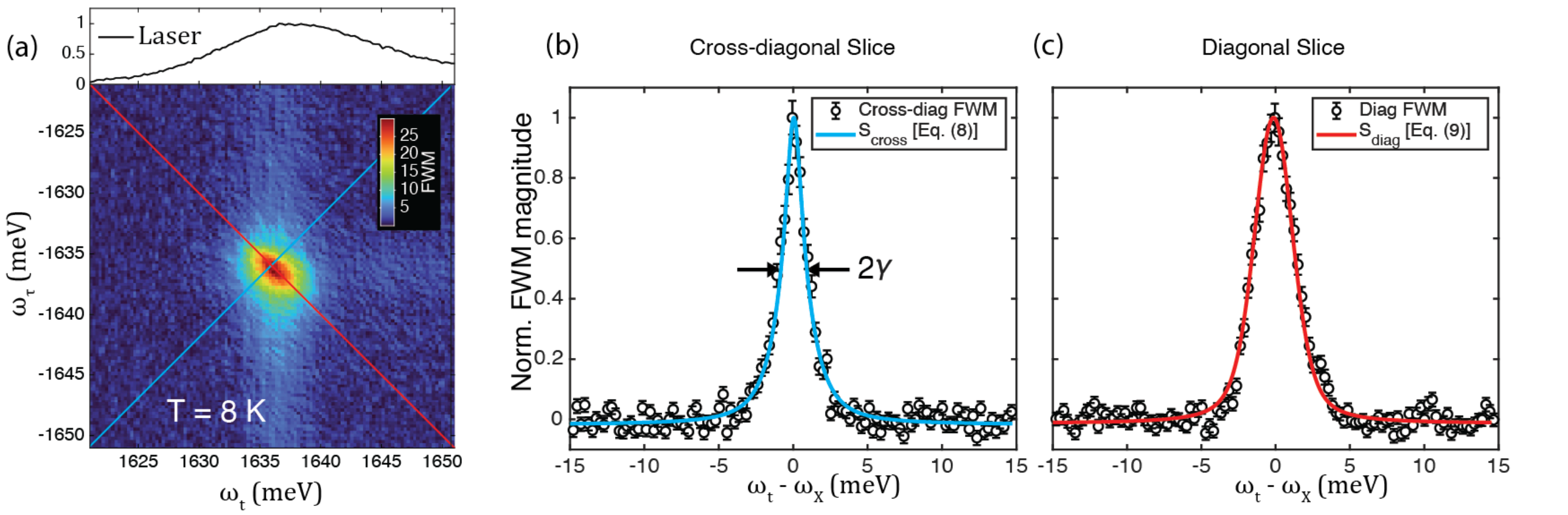

Fig. 3. **Two-dimensional coherent spectrum and extracted lineshape slices of the $MoSe_2$ exciton at 8 K.** (a) Amplitude of the 2DCS spectrum at waiting time $T = 200$ fs, with the measured laser spectrum, reflected off bare $SiO_2$, shown along the top. Blue and red lines mark the cross-diagonal and diagonal slices. (b) Cross-diagonal slice and best fit to Eq. (8), with homogeneous linewidth $2\gamma$. (c) Diagonal slice and best fit to Eq. (9).

Profiling the joint $\chi^2$ surface over $(\sigma,\gamma)$ produces the confidence regions shown in Fig. 4. Relative to PL, their reduced elongation demonstrates improved linewidth separation from the complementary diagonal and cross-diagonal constraints.

To compare the PL and 2DCS confidence regions quantitatively, we fit a quadratic form

$$\Delta\chi^2(\Delta\sigma,\Delta\gamma) \approx a(\Delta\sigma)^2 + 2b(\Delta\sigma)(\Delta\gamma) + c(\Delta\gamma)^2, \tag{11}$$

to each profiled $\Delta\chi^2$ surface near its minimum, yielding the curvature matrix

$$K = \begin{pmatrix} a & b \\ b & c \end{pmatrix} \tag{12}$$

and local covariance matrix $C = K^{-1}$ [24]. From $C$, we extract the Pearson correlation coefficient $\rho_{\sigma\gamma} = C_{\sigma\gamma}/\sqrt{C_{\sigma\sigma}C_{\gamma\gamma}}$, and the axis ratio $r = \sqrt{\lambda_1/\lambda_2}$, where $\lambda_1 \geq \lambda_2$ are the eigenvalues of $C$.

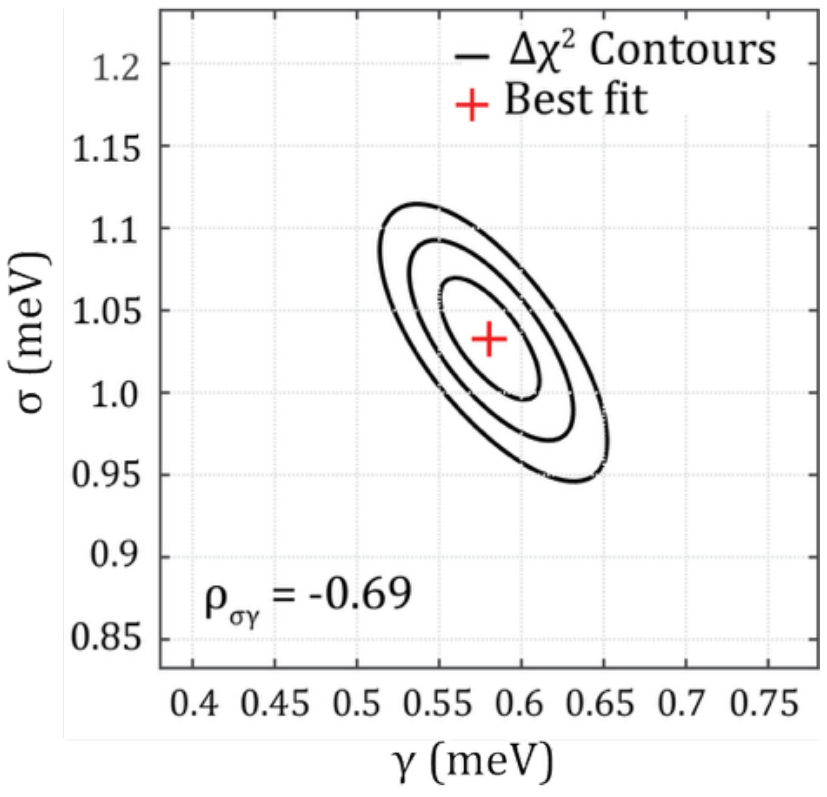

Fig. 4. **Joint 2DCS fitting improves homogeneous-inhomogeneous linewidth separation**. Profiled $\Delta\chi^2_{2DCS}(\sigma,\gamma)$ landscape from Eq. (10). Contours show the 68.27%, 95.45%, and 99.73% joint confidence regions; the red + marks the best fit.

To assess spatial robustness, we repeat the analysis for 2DCS spectra at 12 locations and PL spectra at 36 locations across similar regions of the hBN-encapsulated $MoSe_2$ flake. Hyperspectral PL imaging maintains spatial registration and maps the exciton energy, intensity, and width, enabling reproducible sampling of corresponding regions between measurements [27]. At each location, the full confidence-landscape analysis is performed independently, yielding distributions of $\rho$, $r$, $\sigma$, and $\gamma$.

Fig. 5 summarizes the quantitative comparison. PL gives $|\rho_{\sigma\gamma}| = 0.93 \pm 0.01$ and $r = 5.1 \pm 0.2$, whereas joint 2DCS fitting gives $|\rho_{\sigma\gamma}| = 0.73 \pm 0.03$ and $r = 2.6 \pm 0.2$. Thus, 2DCS produces reduced parameter correlation and less elongated confidence regions across the measured locations.

The best-fit linewidth partitions also differ between the two measurements. Across all locations, reported as the mean $\pm$ standard deviation, PL gives $\sigma_{\text{PL}} = 0.56 \pm 0.07$ meV and $\gamma_{\text{PL}} = 1.00 \pm 0.07$ meV, whereas 2DCS gives $\sigma_{\text{2D}} = 0.91 \pm 0.08$ meV and $\gamma_{\text{2D}} = 0.58 \pm 0.08$ meV. Reduced parameter correlation does not establish that one linewidth partition is universally more accurate because PL and 2DCS probe different observables. PL weights populated emissive states affected by relaxation, localization, and radiative efficiency, whereas 2DCS probes the coherent nonlinear response of optically bright excitons within the excitation bandwidth. The physically relevant partition, therefore, depends on whether the question concerns emission or coherent optical dephasing. In principle, absorption or reflection measurements eliminate the effects of relaxation; transmission experiments are impossible for samples on opaque substrates, and interpretation of reflection spectra can be difficult because of multilayer interference, both of which are the case here. The comparison shows that the PL decomposition is strongly under-constrained, whereas the complementary 2DCS projections determine the two linewidth contributions more independently.

The confidence regions quantify statistical identifiability within the assumed symmetric lineshape models, not model-dependent systematic uncertainty. Exciton–phonon coupling can produce asymmetric PL and diagonal 2DCS lineshapes [23,28], while excitation-induced dephasing and shifts may modify the measured amplitude spectrum [15,18]. Quantitatively separating these effects would require additional measurements and microscopic lineshape models beyond the scope of the present comparison.

In summary, we quantitatively compare homogeneous–inhomogeneous linewidth separation by one-dimensional Voigt analysis and joint fitting of diagonal and cross-diagonal 2DCS slices. The complementary sensitivity of the two 2DCS projections produces confidence regions with reduced parameter correlation and elongation. Although demonstrated using hBN-encapsulated $MoSe_2$ at 8 K, the comparison applies more broadly to systems described by Gaussian inhomogeneous broadening and Lorentzian homogeneous dephasing. The numerical degree of improvement will depend on the relative linewidths, signal-to-noise ratio, spectral resolution, and validity of the assumed lineshape model. These results quantify when one-dimensional Voigt analysis leaves the linewidth decomposition underconstrained and when additional spectroscopic constraints are beneficial.

**Funding.** A.A. and S.T.C. acknowledge funding from NSF Grant 2004286. J.R.H. acknowledges support from AFOSR under award no. FA9550-25RYCOR006.

**Disclosures.** STC is a co-founder of MONSTR Sense Technologies LLC, which manufactures co-linear MDCS Spectrometers.

**Data availability.** The data that support the findings of this study are available from the corresponding author upon reasonable request.

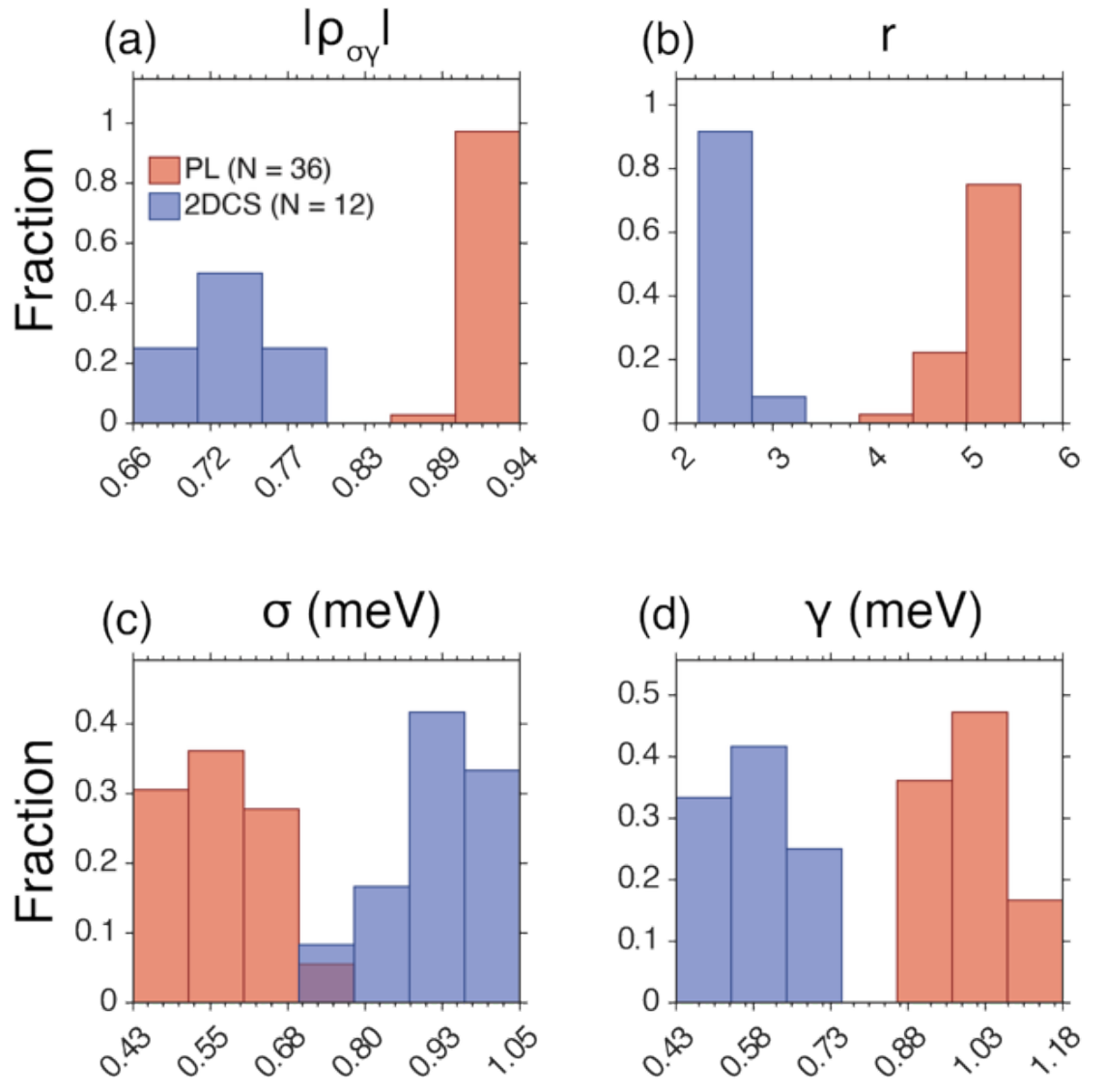

Fig. 5. **Quantitative comparison of linewidth separation by PL and 2DCS.** Histograms of (a) the magnitude of the Pearson correlation coefficient $|\rho_{\sigma\gamma}|$, (b) confidence-ellipse axis ratio $r$, (c) best-fit $\sigma$, and (d) best-fit $\gamma$. Relative to PL, joint 2DCS fitting yields lower $|\rho_{\sigma\gamma}|$ and $r$, indicating improved parameter separability across the measured locations.

Extended references: